\documentclass[conference]{IEEEtran}
\usepackage{graphicx}
\usepackage{epstopdf}
\usepackage[latin1]{inputenc}
\usepackage{amsmath,amssymb,amscd,latexsym,dsfont}
\usepackage[english]{babel}
\usepackage{tabularx,cite}
\usepackage{graphicx}
\usepackage{algpseudocode}
\usepackage{algorithm}
\usepackage{algorithmicx}
\usepackage{float}
\usepackage{multicol}
\usepackage{psfrag}
\usepackage{comment,psfrag,subfigure,enumerate}
\usepackage{color}
\usepackage{amsthm}
\usepackage{graphicx}
\usepackage{graphicx}
\usepackage{epstopdf}
\usepackage[latin1]{inputenc}
\usepackage{amsmath,amssymb,amscd,latexsym,dsfont}
\usepackage[english]{babel}
\usepackage{tabularx,cite}
\usepackage{graphicx}
\usepackage{algpseudocode}
\usepackage{algorithm}
\usepackage{algorithmicx}
\usepackage{float}
\usepackage{multicol}
\usepackage{mathtools}
\usepackage{psfrag}
\usepackage{comment,psfrag,subfigure,enumerate}
\usepackage{color}
\usepackage{amsthm}
\ifCLASSINFOpdf
\else
\fi
\begin{document}
\IEEEoverridecommandlockouts
\IEEEpubid{\makebox[\columnwidth]{978-1-4799-5863-4/14/\$31.00 \copyright 2014 IEEE \hfill} \hspace{\columnsep}\makebox[\columnwidth]{ }}
\title{On the Performance of Millimeter Wave-based RF-FSO Links with HARQ Feedback}

\author{
    \IEEEauthorblockN{Behrooz Makki\IEEEauthorrefmark{1}, Tommy Svensson\IEEEauthorrefmark{1}, and Mohamed-Slim Alouini\IEEEauthorrefmark{2}}
    \IEEEauthorblockA{\IEEEauthorrefmark{1}Chalmers University of Technology, Gothenburg, Sweden, \{behrooz.makki, tommy.svensson\}@chalmers.se}
    \IEEEauthorblockA{\IEEEauthorrefmark{2} King Abdullah University of Science and Technology (KAUST), Thuwal, Saudi Arabia, slim.alouini@kaust.edu.sa}
    %\thanks{This work was supported in part by the Swedish Governmental Agency for Innovation Systems (VINNOVA) within the VINN Excellence Center Chase.}
}

%Telephone: (800) 555--1212, Fax: (888) 555--1212}


\maketitle
%\onecolumn
\vspace{-0mm}
\begin{abstract}
This paper studies the performance of hybrid radio-frequency (RF) and free-space optical (FSO) links in the cases with and without hybrid automatic repeat request (HARQ). Considering millimeter wave (mmwave) characteristics in the RF link and pointing errors in the FSO link, we derive closed-form expressions for the message decoding probabilities as well as the throughput and the outage probability of the RF-FSO setups. We also evaluate the effect of various parameters such as power amplifiers efficiency, different transmission techniques in the FSO link, pointing errors in the FSO link as well as different coherence times/symbol rates of the RF and the FSO links on the throughput and outage probability. The results show the efficiency of the RF-FSO links in different conditions. Moreover, the HARQ can effectively improve the outage probability/energy efficiency, and compensate the effect of hardware impairments in RF-FSO links.
\end{abstract}
%Finally, we derive closed-form expressions for the maximum achievable throughput of the return-link satellite systems using optimal schedulers.

%Then, compared to open-loop communication setups, the implementation of power-adaptive ARQ reduces the average power by ? and ? dB, if a maximum of 2 and 3 retransmissions is utilized, respectively.


% \fi
\IEEEpeerreviewmaketitle
\vspace{-2mm}
\section{Introduction}
The demand for high data rates raises the question of spectrum availability. %Particularly, reaching rates of 10 Gbps or more is challenging due to the major limitation factor, scarce spectrum resources.
Thus, for future communication systems, more spectral resources are mandatory. Of the many popular solutions, free-space optical (FSO) communication systems have gained significant research attention as effective
means of transferring data at high rates over short distances. In the radio frequency (RF) domain, on the other hand, it is mainly concentrated on millimeter wave (mmwave) communication as a key enabler to obtain sufficiently large bandwidths so that it is possible to achieve data rates comparable to those in the FSO links. In this perspective, the link reliability and the service availability can be considerably improved via the combination of FSO and mmwave-based RF links. This is particularly because both the FSO and the mmwave-based RF links are highly susceptible to atmospheric effects. However, the good point is that these links are complementary because the RF (resp. the FSO) signal is severely attenuated by rain (resp. fog/clouds) while the FSO (resp. the RF) signal is not.

%The next generation of communication networks must provide high-rate reliable data streams. To address these demands, a combination of different techniques are considered among which free-space optical (FSO) communication is very promising  \cite{6331134,6932439,6887284}. FSO systems provide fiber-like data rates through the atmosphere using lasers or light emitting diodes (LEDs). Thus, the FSO can be used for a wide range of applications such as last-mile access, fiber back-up, back-haul for wireless cellular networks, and disaster recovery. However, such links are highly susceptible to atmospheric effects and, consequently, are unreliable. An efficient method to improve the reliability in FSO systems is to rely on an additional radio-frequency (RF) link to create a hybrid RF-FSO communication system.

%Typically, to achieve data rates comparable to those in the FSO link, a millimeter wavelength carrier is selected for the RF link. As a result, the RF link is also subject to atmospheric effects such as rain and scintillation.

The performance of RF-FSO systems is studied in different papers, e.g., \cite{6887284,1399401,6364576}, where the RF and the FSO links are considered as separate links and the RF link acts as a backup when the FSO link is down. On the other hand, in \cite{6503564,5342330,4411336,4348339,5351671} the RF and the FSO links are combined to improve the system performance. Moreover, the implementation of hybrid automatic repeat request (HARQ) in RF-based (resp. FSO-based) systems is investigated in, e.g., \cite{throughputdef,a01661837,MIMOARQkhodemun,tuninetti2011,6235077} (resp. \cite{6168189,6139812,5380093,5557647,5680280}), while the HARQ-based RF-FSO systems have been rarely studied, e.g., \cite{5427418,6692504,RFFSOglobecom}.

To have realistic insights about the performance of RF-FSO links, it is necessary to take the non-ideal link properties into account. Particularly, considering the FSO link,  thermal expansion, dynamic wind loads, and weak earthquakes result in the building sway phenomenon that causes vibration of the transmitter beam leading to a misalignment between the FSO transmitter and receiver known as pointing error\footnote{In general, the mmwave-based RF links also suffer from pointing errors in cases with narrow beamforming and mobility. However, here we focus on static links (e.g., wireless backhauling), in which pointing errors of the RF link are negligible.}. The pointing error may lead to significant performance degradation and is a serious issue in urban areas, where the FSO equipments are placed on high-rise buildings \cite{7192727,5936940}. In the RF link, on the other hand, the power amplifiers (PAs) efficiency is the main hardware problem affecting the system performance \cite{7104158,4160747}. These are the main motivations for our paper analyzing the performance of HARQ-based RF-FSO links with pointing errors and imperfect RF PAs.

\begin{figure}
\centering
  \includegraphics[width=0.99\columnwidth]{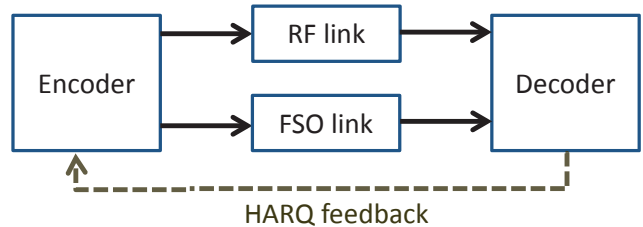}\\\vspace{-2mm}
\caption{Channel model. The data is jointly transmitted by the RF and the FSO links and, in each round of HARQ, the receiver decodes the data based on all received signals.}\vspace{-2mm}\label{figure111}
\end{figure}
\vspace{-0mm}
In this paper, we study the data transmission efficiency of RF-FSO systems from an information theoretic point of view. The contributions of this paper are twofold. 1) Taking the pointing errors in the FSO link and the mmwave characteristics of the RF link into account, we derive closed-form expressions for the message decoding probabilities as well as the system throughput and outage probability. Our results are obtained in different cases with and without HARQ. Particularly, we show the HARQ as an effective technique to compensate the non-ideal properties of the RF-FSO links and improve the hybrid link reliability. Then, 2) we analyze the effect of various parameters such as the FSO link pointing errors, the PAs efficiency, different heterodyne detection- and intensity modulation with direct detection (IM/DD)-based data transmission techniques of the FSO link, and different symbol rates/coherence times of the RF and FSO links on the throughput/outage probability.

In contrast to \cite{6887284,1399401,6364576}, we consider joint data transmission/reception in the RF and FSO links. Moreover, our paper is different from \cite{6503564,5342330,7104158,6235077,4160747,4411336,4348339,5351671,5936940,throughputdef,a01661837,MIMOARQkhodemun,tuninetti2011,6168189,6139812,5380093,5557647,5680280,5427418,6692504,7192727,4432329,RFFSOglobecom} because we derive new analytical/numerical results on the message decoding probabilities, and outage probability/throughput of HARQ-based RF-FSO links which have not been presented before.  The differences in the problem formulation and the channel model makes our analytical/numerical results and conclusions completely different from the ones in the literature, e.g., \cite{6887284,1399401,6364576,6503564,5342330,7104158,6235077,4160747,4411336,4348339,5351671,5936940,throughputdef,a01661837,MIMOARQkhodemun,tuninetti2011,6168189,6139812,5380093,5557647,5680280,5427418,6692504,7192727,4432329,RFFSOglobecom}.

The numerical and the analytical results show the efficiency of RF-FSO links in different conditions. Moreover, the HARQ protocols can effectively improve the outage probability/energy efficiency, and compensate the effect of hardware impairments in RF-FSO links. For instance, with the common parameter settings of the RF-FSO links, outage probability $10^{-2}$ and code rate $1$ nats-per-channel-use (npcu), the implementation of HARQ with a maximum of $3$ retransmissions reduces the required power by $8$ dB, compared to the cases with open-loop communication.
\vspace{-0mm}
\section{System Model}
Consider a joint RF-FSO system, as demonstrated in Fig. 1. The data sequence is encoded
into parallel FSO and RF bit streams.
%The FSO link employs intensity modulation and direct detection while the RF link modulates the encoded bits and up-converts the baseband signal to a millimeter wavelength, in the range of $30-300$ GHz, RF carrier frequency.
Then, the FSO and the RF signals are simultaneously sent to the receiver. At the receiver, the received RF (FSO) signal is down-converted to baseband (resp. collected by an aperture and converted to an electrical signal via photo-detection) and the signals are sent to the decoder which decodes the received signals jointly.
%In this way, as seen in the following, the diversity increases by the joint data transmission of the RF and FSO links, and one link can compensate the poor performance of the other link experiencing severe atmospheric effects.

We denote the instantaneous realizations of the fading coefficient of the RF link and the turbulence coefficient of the FSO link in time slot $i$ by $H_{\text{RF},i}$ and $H_{\text{FSO},i}$, respectively, and for simplicity we refer to both of them as channel coefficients. These channel coefficients are assumed to be known at the receiver which is an acceptable assumption in block-fading conditions  \cite{throughputdef,a01661837,MIMOARQkhodemun,tuninetti2011,6168189,6139812,7192727,6235077}. Also, we define $G_{\text{RF},i}=|H_{\text{RF},i}|^2,$ $G_{\text{FSO},i}=|H_{\text{FSO},i}|^2$ which are referred to as channel gain realizations in the following. We then assume no channel state information (CSI) feedback to the transmitter, except for the HARQ feedback bits. The feedback channel can be an RF, an FSO or an RF-FSO link, and is supposed to be error- and delay-free. Finally, we assume perfect synchronization between the links in harmony with, e.g., \cite{6503564,5342330,4411336,4348339,5351671,5427418,6692504}.

As the most promising HARQ approach leading to highest throughput/lowest outage probability \cite{throughputdef,MIMOARQkhodemun,a01661837,tuninetti2011}, we consider the incremental redundancy (INR) HARQ with a maximum of $M$ retransmissions, i.e., the message is retransmitted a maximum of $M$ times. Using INR HARQ, $K$ information nats are encoded into a \emph{parent} codeword of length $ML$ channel uses. The parent codeword is then divided into $M$ sub-codewords of length $L$ channel uses which are sent in the successive transmission rounds. Thus, the equivalent data rate, i.e., the code rate, at the end of round $m$ is $\frac{K}{mL}=\frac{R}{m}$ npcu where $R=\frac{K}{L}$ denotes the initial code rate. In each round, the receiver combines all received sub-codewords to decode the message. The retransmission continues until the message is correctly decoded or the maximum permitted transmission round is reached. Note that setting $M=1$ represents the cases without HARQ, i.e., open-loop communication.

The FSO link is assumed to follow a unified Gamma-Gamma fading distribution with pointing errors for which the probability distribution function (PDF) of the channel gain is given by
\begin{align}\label{eq:Eqpointerrorpdf}
f_{G_{\text{FSO}}}(x)=\frac{\xi^2}{rx\Gamma(\alpha)\Gamma(\beta)}\mathcal{G}_{1,3}^{3,0}\Bigg(h\alpha\beta\left(\frac{x}{\mu_r}\right)^\frac{1}{r}\Bigg|_{\xi^2,\alpha,\beta}^{\xi^2+1}\Bigg).
\end{align}
This is a well-established model for pointing errors in FSO links, e.g., \cite{7192727,5936940}. Also, $\Gamma(\cdot)$ denotes the Gamma function. Then, $r$ represents the parameter specifying the detection technique type, where $r=1$ accounts for heterodyne detection and $r=2$ represents IM/DD technique. Also, $h=\frac{\xi^2}{\xi^2+1}$ denotes the ratio between the equivalent beam radius at the receiver and the pointing error displacement standard deviation (jitter) at the receiver \cite{5936940} (i.e. for negligible pointing errors, $\xi\to\infty$). Then, $\alpha$ and $\beta$ are the fading/scintillation parameters related to the atmospheric turbulence conditions, and $\mathcal{G}_{\cdot,\cdot}^{\cdot,\cdot}(\cdot)$ is the Meijer's G function as defined in \cite[Eq.(9.301)]{MeijerGbood}. Finally,  $\mu_r$ stands for the average electrical signal-to-noise ratio (SNR) where $\mu_1=\mu_\text{heterodyne}=E[{G_{\text{FSO}}}]$ and $\mu_2=\mu_\text{IM/DD}=\frac{E[{G_{\text{FSO}}}]\alpha\beta\xi^2(\xi^2+2)}{(\alpha+1)(\beta+1)(\xi^2+1)^2}$ \cite{7192727}. Finally, we denote the transmission power of the FSO link by $P_\text{FSO}.$

While the modeling of the mmwave-based links is well known for line-of-sight wireless backhaul links, it is still an ongoing research topic  for non-line-of-sight conditions \cite{mmmagic}. Particularly, different measurement setups have emphasized the non-/near-line-of-sight propagation and the non-ideal hardware as two key challenges of such links. Since we target a hybrid mmwave-based RF-FSO link, we consider Rician channel model for the RF link, which is an appropriate model for near line-of-sight conditions. With a Rician model, the channel amplitude $\sqrt{G_\text{RF}}$ and gain ${G_\text{RF}}$, respectively, follow the PDFs
\begin{align}\label{eq:eqRicianpdf}
\tilde f_{\text{RF}}(x)=\frac{x}{\omega}e^{-\frac{(x^2+\nu^2)}{2\omega^2}}I_0\left(\frac{x\nu}{\omega^2}\right),
\end{align}
and $f_{G_\text{RF}}(x)=\frac{1}{2\sqrt{x}}\tilde f_{\text{RF}}(\sqrt{x})$, where $\nu$ and $\omega$ denote the fading parameters and $I_0$ is the zero-th order modified Bessel function of the first kind.

Finally, to take the non-ideal hardware into account, we consider the state-of-the-art model for the PA efficiency in the RF link which is given by
\cite{7104158,4160747}
\begin{align}\label{eq:ampmodeldaniel}
&\frac{P_\text{RF}}{P_\text{RF}^\text{cons}}=\epsilon\left(\frac{P_\text{RF}}{P_\text{RF}^\text{max}}\right)^\vartheta\Rightarrow  P_\text{RF}=\sqrt[1-\vartheta]{\frac{\epsilon P_\text{RF}^\text{cons}}{(P_\text{RF}^\text{max})^\vartheta}}.
\end{align}
Here, $P_\text{RF}, P_\text{RF}^\text{max}$ and $P_\text{RF}^\text{cons}$ are the output, the maximum output and the consumed power of the PA, respectively, $\epsilon\in [0,1]$ denotes the maximum power efficiency achieved at $P_\text{RF}=P_\text{RF}^\text{max}$ and $\vartheta\in [0,1]$ is a parameter depending on the PA classes.

\section{Analytical results}
As shown in \cite{throughputdef,a01661837,MIMOARQkhodemun,tuninetti2011,6235077}, for different channel models the throughput of HARQ protocols can be written as
\begin{align}\label{eq:eqeta1}
\eta=\frac{R(1-\phi_M)}{1+\sum_{m=1}^{M-1}{\phi_m}},
\end{align}
%and
%\vspace{-3mm}
%\begin{align}\label{eq:eqoutprob1}
%\Pr(\text{Outage})=\Pr\left(W_M\le \frac{R}{M}\right),
%\end{align}
%%
%%\begin{align}\label{eq:eqfunc}
%%\eta=\text{Function}(\Pr(W_m\le \frac{R}{m}),\forall m=1,\ldots,M)
%%\end{align}
%respectively,
where
%$W_m$ is the accumulated mutual information (AMI) at the end of round $m$. Also,
$\phi_m=\Pr(W_m\le \frac{R}{m})$ denotes the probability that the data is not correctly decoded up to the end of the $m$-th round with $W_m$ being the accumulated mutual information (AMI) at the end of round $m$.  Also, the outage probability is given by $\Pr(\text{Outage})=\phi_M=\Pr(W_M\le \frac{R}{M})$. Thus, to analyze the throughput and outage probability, the key point is to determine the
%AMIs as functions of channel realization(s) and find their corresponding cumulative distribution functions (CDFs)\footnote{The CDF and the probability distribution function (PDF) of a random variable $X$ are denoted by $F_X(.)$ and $f_X(.),$ respectively.} $F_{W_{m}},m=1,\ldots,M$
probabilities $\phi_m,m=1,\ldots,M$. Then, having the probabilities, the considered performance metrics are obtained.

\begin{figure}
\centering
  \includegraphics[width=0.99\columnwidth]{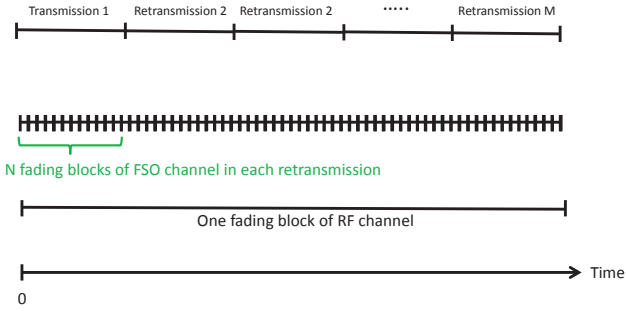}\\\vspace{-3mm}
\caption{Time scales. The RF link remains constant in the retransmissions (quasi-static channel \cite{a01661837,MIMOARQkhodemun}) while in each retransmission round of HARQ $N$ different channel realizations are experienced in the FSO link.}\vspace{-3mm}\label{figure111}
\end{figure}

In RF-FSO systems, it was demonstrated by, e.g., \cite{5342330,598416} and the references therein, that the RF link experiences very slow variations and the coherence time of the RF link is in the order of $10^{2}-10^{3}$ times larger than the coherence time of the FSO link. For this reason, we consider the setup as illustrated in Fig. 2 where the RF link remains constant in the retransmissions (quasi-static channel \cite{a01661837,MIMOARQkhodemun}) while in each retransmission round of HARQ $N$ different channel realizations are experienced in the FSO link.
%However, note that this is not a necessary condition because 1) the same analysis holds for the cases with shorter coherence time of the RF link, compared to the coherence time of the FSO link and 2) as seen in Section III.B, we can derive the results in the cases with few, possibly 1, channel realizations of the FSO link during the packet transmission.
In this way, considering (\ref{eq:ampmodeldaniel}) and Fig. 2, we can use the results of \cite[Chapter 7]{ELGAMAL} to find the probability $\phi_m,\forall m=1,\ldots,M,$ as
\vspace{-1mm}
\begin{align}\label{eq:eqWm1}
&\phi_m
=\Pr\left(\log\left(1+\sqrt[1-\vartheta]{\frac{\epsilon P_\text{RF}^\text{cons}}{(P_\text{RF}^\text{max})^\vartheta}}G_\text{RF}\right)+\mathcal{Y}_{r,(m,N)}\le\frac{R}{M}\right),
\nonumber\\&\mathcal{Y}_{r,(m,N)}\doteq\frac{\psi}{mN}\sum_{j=1}^{m}{\sum_{i=1}^N{\log\left(1+c_rP_\text{FSO}G_{\text{FSO}, 1+(j-1)i}\right)}}.
\end{align}
Here,
%$P_\text{RF}$ and $P_\text{FSO}$ are, respectively, the transmission powers in the RF and FSO links. Also, $G_\text{RF}$ and $G_{\text{FSO},j}$'s denote the channel gain realizations of the RF and the FSO links in different retransmission rounds, respectively. Then,
$\psi$ represents the relative symbol rates of the RF and FSO links which is a design parameter selected by the network designer.
%With no loss of generality, we set $\psi=1$ in the following, while the results can be easily extended to the cases with different values of $\psi$.
Also, (\ref{eq:eqWm1}) is based on the fact that the achievable rate of an FSO link is given by $\log(1+c_rx)$ with $x$ being the instantaneous received SNR and $c_r$ denoting a constant term such that $c_r = 1$ for heterodyne detection and $c_r =\frac{e}{2\pi}$ for IM/DD \cite{7192727,alouinicapacityakhar}, \cite[Eq. 7.43]{fsobookalouini}, \cite[Eq. 26]{5238736}\footnote{In \cite{5238736}, $\log(1+c\text{SNR})$ is proved as a tight lower bound on the capacity in the cases with an average power constraint. Then, \cite{7192727,alouinicapacityakhar} show that the formula of the kind $\log(1+c\text{SNR})$ is an asymptotically tight lower bound on the achievable rates for the cases with an average power constraint, a peak power constraint, as well as combined peak and average power constraints.}.

%In the following, we set $c=1$ which corresponds to the cases with heterodyne detection. In the meantime, setting $c =\frac{e}{2\pi}$, it is straightforward to extend the results to the cases with IM/DD. Finally, as the noise variances at the receiver are set to 1, we define $10\log_{10}P, P=P_\text{RF}+P_\text{FSO},$ as the SNR.
%\vspace{-3mm}
%\subsection{Performance Analysis in the Cases with Considerably Different Coherence Times for the RF and FSO Links}
%Here, we consider the cases where the coherence times of the RF and FSO links are considerably different. Motivated by, e.g., \cite{5342330,1142964,598416,598420}, we concentrate on the cases with shorter coherence time of the FSO link, compared to the RF link. Meanwhile, the same analysis is valid, if the RF link experiences shorter coherence time than the FSO link.

Considering different values of $N$, there is no closed-form expression for (\ref{eq:eqWm1}). Thus, we use the central limit theorem to approximate $\mathcal{Y}_{r,(m,N)}$ by the Gaussian random variable $\mathcal{Z}\sim\mathcal{N}(\mu,\frac{1}{mN}\sigma^2)$. Here, $\mu$ and $\sigma^2$ are the mean and variance determined based on the FSO link channel condition. Particularly, considering the Gamma-Gamma PDF (\ref{eq:Eqpointerrorpdf}) and denoting the expectation operator by $E\{\cdot\},$ we have
\begin{align}\label{eq:eqmeanerrorpoint}
&\mu=\psi E\{\log(1+c_rP_\text{FSO}G_\text{FSO})\}\nonumber\\&
=\frac{\xi^2\psi}{rx\Gamma(\alpha)\Gamma(\beta)}\times\nonumber\\&\int_0^\infty{\log(1+c_rP_\text{FSO}x)\mathcal{G}_{1,3}^{3,0}\Bigg(h\alpha\beta\left(\frac{x}{\mu_r}\right)^\frac{1}{r}\Bigg|_{\xi^2,\alpha,\beta}^{\xi^2+1}\Bigg)\text{d}x},
\end{align}
%Reviewing the literature and depending on the channel condition, the FSO link is commonly considered to follow exponential, log-normal or Gamma-Gamma distributions, e.g., \cite{5342330,6168189,FSObook}. For the exponential distribution of the FSO link, i.e., $f_{G_\text{FSO}}(x)=\lambda e^{-\lambda x}$ with $\lambda$ being the long-term channel coefficient, we have
%\begin{align}\label{eq:mueq}
%\mu&=E\{\log(1+P_\text{FSO}G_\text{FSO})\}=\int_0^\infty{f_{{G_{\text{FSO}}}}(x)\log(1+P_\text{FSO}x)\text{d}x}\nonumber\\&\mathop  = \limits^{(a)} P_\text{FSO}\int_0^\infty{\frac{1-F_{G_{\text{FSO}}}(x)}{1+P_\text{FSO}x}\text{d}x}=-e^{\frac{\lambda }{P_\text{FSO}}}\text{Ei}\left(-{\frac{\lambda }{P_\text{FSO}}}\right)
%\end{align}
and $\sigma^2=\rho^2-\mu^2$ with
\begin{align}\label{eq:sigmaeq}
&\rho^2=\psi^2E\{\log(1+c_rP_\text{FSO}G_\text{FSO})^2\}
=\frac{\xi^2\psi^2}{rx\Gamma(\alpha)\Gamma(\beta)}\times\nonumber\\&\int_0^\infty{\log(1+c_rP_\text{FSO}x)^2\mathcal{G}_{1,3}^{3,0}\Bigg(h\alpha\beta\left(\frac{x}{\mu_r}\right)^\frac{1}{r}\Bigg|_{\xi^2,\alpha,\beta}^{\xi^2+1}\Bigg)\text{d}x}
\end{align}
which can be found numerically, because they are one-dimensional integrations.

Having $\mu$ and $\sigma^2$, we find the probabilities $\phi_m,\forall m,$ as follows. Considering Rician fading conditions (\ref{eq:eqRicianpdf}) for the RF link, (\ref{eq:eqWm1}) is rephrased as
\begin{align}
&\phi_m=
\int_0^{d_m}{\frac{\tilde f_{\text{RF}}(\sqrt{x})}{2\sqrt{x}}\times}\nonumber\\&
\Pr\left(\mathcal{Y}_{r,(m,N)}\le \frac{R}{m}-\log\left(1+\sqrt[1-\vartheta]{\frac{\epsilon P_\text{RF}^\text{cons}}{(P_\text{RF}^\text{max})^\vartheta}}x\right)\right)\text{d}x=
\nonumber\\&\int_0^{d_m}{\frac{\tilde f_{\text{RF}}(\sqrt{x})}{2\sqrt{x}}Q\bigg(\frac{\sqrt{mN}\left(\log\left(1+\sqrt[1-\vartheta]{\frac{\epsilon P_\text{RF}^\text{cons}}{(P_\text{RF}^\text{max})^\vartheta}}x\right)+\mu-\frac{R}{m}\right)}{\sigma}\bigg)\text{d}x}
\nonumber\\
&
\mathop  = \limits^{(a)}\int_0^{d_m^2}{{\tilde f_{\text{RF}}(u)}\times}
\nonumber\\&\,\,\,\,\,\,\,\,\,\,\,\,\,\,\,\,\,\,\,\,\,\,\,\,\,Q\left(\frac{\sqrt{mN}\left(\log\left(1+\sqrt[1-\vartheta]{\frac{\epsilon P_\text{RF}^\text{cons}}{(P_\text{RF}^\text{max})^\vartheta}}u^2\right)+\mu-\frac{R}{m}\right)}{\sigma}\right)\text{d}u
\nonumber\\&
\mathop  \simeq \limits^{(b)}\int_0^{d_m^2}{{\tilde f_{\text{RF}}(u)}V_{\tau_m,\lambda_m}(u)\text{d}u}
\nonumber\\&
=\int_0^{ a_m}{{\tilde f_{\text{RF}}(u)}\text{d}u}+\int_{ a_m}^{ b_m}{\tilde f_{\text{RF}}(u)\left(\frac{1}{2}-\lambda_m(u-\tau_m)\right)\text{d}u}
\nonumber
\end{align}
\begin{align}\label{eq:eqbessel0}
&
\mathop  \simeq \limits^{(c)}{\tilde F_{\text{RF}}( a_m)}+\left(\frac{1}{2}+\lambda_m\tau_m\right)\left({\tilde F_{\text{RF}}( b_m)}-{\tilde F_{\text{RF}}( a_m)}\right)\nonumber\\&-\lambda_m\bigg( b_m\tilde F_{\text{RF}}( b_m)- a_m\tilde F_{\text{RF}}( a_m)\nonumber\\&\,\,\,\,\,\,\,\,\,\,\,\,\,\,\,\,\,\,\,\,\,\,\,\,-\left( b_m- a_m\right)\tilde F_{\text{RF}}\left(\frac{ a_m+ b_m}{2}\right)\bigg),
\nonumber\\&
d_m\doteq\frac{e^{\frac{R}{m}}-1}{\sqrt[1-\vartheta]{\frac{\epsilon P_\text{RF}^\text{cons}}{(P_\text{RF}^\text{max})^\vartheta}}},
\tau_m\doteq\frac{e^{\frac{R}{m}-\mu}-1}{\sqrt[1-\vartheta]{\frac{\epsilon P_\text{RF}^\text{cons}}{(P_\text{RF}^\text{max})^\vartheta}}},
\nonumber\\&
\lambda_m\doteq\sqrt{\frac{2mN\left(e^{-(\frac{R}{m}-\mu)}-e^{-2(\frac{R}{m}-\mu)}\right)}{\pi\sigma^2}\sqrt[1-\vartheta]{\frac{\epsilon P_\text{RF}^\text{cons}}{(P_\text{RF}^\text{max})^\vartheta}}},
\nonumber\\&
a_m\doteq{\max\left(0, \tau_m-\frac{1}{2 \lambda_m}\right)},
b_m\doteq{\min\left( \tau_m+\frac{1}{2 \lambda_m},d_m\right)}.
\end{align}
Here, $(a)$ is obtained by variable transform $u=\sqrt{x}.$ Then, $(b)$ comes from the linear approximation technique $Q\left(\frac{\sqrt{mN}\left(\log\left(1+\sqrt[1-\vartheta]{\frac{\epsilon P_\text{RF}^\text{cons}}{(P_\text{RF}^\text{max})^\vartheta}}u^2\right)+\mu-\frac{R}{m}\right)}{\sigma}\right)\simeq V_{\tau_m,\lambda_m}(x)$ with
\begin{align}\label{eq:linearizationnik2}
&V_{\tau_m,\lambda_m}(x)= \left\{\begin{matrix}
1 & x< a_m,\\
\frac{1}{2}-\lambda_m(x-\tau_m) & a_m\le x\le b_m,\\
0 & x> b_m,
\end{matrix}\right.\vspace{-2mm}
\end{align}
where $\lambda_m$ defined in (\ref{eq:eqbessel0}) is obtained by taking the derivative of $Q\left(\frac{\sqrt{mN}\left(\log\left(1+\sqrt[1-\vartheta]{\frac{\epsilon P_\text{RF}^\text{cons}}{(P_\text{RF}^\text{max})^\vartheta}}u^2\right)+\mu-\frac{R}{m}\right)}{\sigma}\right)$ at point $u=\tau_m.$ Then, $(c)$ is obtained by the first order Riemann integral approximation $\int_{x_1}^{x_2}f(x)\text{d}x\simeq(x_2-x_1)f(\frac{x_1+x_2}{2})$. Also, $\tilde F_\text{RF}(x)=1-Q_{\mathcal{M}}\left(\frac{\nu}{\omega},\frac{x}{\omega}\right)$ is the CDF of the Rician variable (\ref{eq:eqRicianpdf}) with $Q_\mathcal{M}(\cdot,\cdot)$ being the Marcum $Q$ function.

Finally, it is interesting to note that $\lim_{N\to\infty}\frac{1}{mN}\sum_{j=1}^{m}{\sum_{i=1}^N{\log(1+c_rP_\text{FSO}G_{\text{FSO}, 1+(j-1)i})}}= E\{\log(1+c_rP_\text{FSO}G_{\text{FSO}})\}, \forall m.$ Intuitively, this means that for asymptotically large values of $N$, i.e., significantly shorter coherence time of the FSO link compared to the one in the RF link, the AMI of the FSO link converges to its ergodic capacity $E\{\log(1+c_rP_\text{FSO}G_{\text{FSO}})\}$. Thus, in this case the RF-FSO link is mapped to an equivalent mmwave-based RF link in which successful decoding of the rate equal to the ergodic capacity of the FSO link is always guaranteed. Also, as a second-order approximation, the probabilities $\phi_m,\forall m=1,\ldots,M,$ are approximated as
\begin{align}\label{eq:eqasymrffso}
\phi_m&\simeq\Pr\left(\log\left(1+\sqrt[1-\vartheta]{\frac{\epsilon P_\text{RF}^\text{cons}}{(P_\text{RF}^\text{max})^\vartheta}}G_\text{RF}\right)\le \frac{R}{m}-\mu\right)\nonumber\\&
=1-Q_{\mathcal{M}}\left(\frac{\nu}{\omega},\frac{e^{\frac{R}{m}-\mu}-1}{\omega\sqrt[1-\vartheta]{\frac{\epsilon P_\text{RF}^\text{cons}}{(P_\text{RF}^\text{max})^\vartheta}}}\right).
\end{align}
Using (\ref{eq:eqbessel0})-(\ref{eq:eqasymrffso}), one can find the probabilities $\phi_m, \forall m=1,\ldots,M,$ and, consequently, the throughput and the outage probability of the RF-FSO system. In Section IV, we verify the accuracy of our derived analytical results and investigate the effect of different parameters such as the PAs efficiency, the FSO link pointing errors, different symbol rates/coherence times of the RF and FSO links and different FSO-based data transmission techniques on the throughput/outage probability of RF-FSO systems.

\section{Numerical Results}
In all figures, we set $\alpha=4.3939, \beta=2.5636$ which correspond to Rytov
variance 1 of the FSO link in the cases with no pointing error \cite{6168189}. Also, the parameters of Rician RF PDF in (\ref{eq:eqRicianpdf}) are set to $\omega=0.7036, \nu=0.0995,$ leading to unit mean and variance of the channel gain distribution $f_{G_\text{RF}}(x)$. Finally, SNR is defined as $10\log_{10} P$ with $P=P_\text{FSO}+P_\text{RF}^\text{cons},$ and we set $P_\text{FSO}=P_\text{RF}^\text{cons}$ in all figures. Then, using (\ref{eq:ampmodeldaniel}), we can find the transmission power of the RF link $P_\text{RF}$ for every given $P_\text{RF}^\text{cons}.$
%Finally, the PA parameters are set to $P^\text{max}=26 \text{ dB},$ $\vartheta=0.5$ and $\epsilon=0.65$, unless otherwise stated.

In Fig. 3, we consider an ideal PA, corresponding to $\epsilon=1, \vartheta=0, P_\text{RF}^\text{max}=\infty$ in (\ref{eq:ampmodeldaniel}), while the effect of imperfect PAs is studied in Figs. 4-6. Particularly, Fig. 3 verifies the accuracy of the approximation techniques (\ref{eq:eqbessel0}) and (\ref{eq:eqasymrffso}), and investigates the outage probability in the cases without ($M=1$) and with HARQ ($M=3$). Here, the results are obtained for the heterodyne detection technique and we set $ R=1 \text{ npcu}, N=100, \xi=0.9, \psi=0.03$.
%Moreover, Fig. 4 studies the outage probability for different values of the relative symbol rates of the links $\psi,$ IM/DD detection technique and $ R=3 \text{ npcu}, N=100, \xi=0.9.$

In Fig. 4, we investigate the effect of imperfect PAs on the system throughput (\ref{eq:eqeta1}). Particularly, considering heterodyne detection, $ R=0.5 \text{ npcu}, N=100, \xi=0.9,$ and $\psi=0.03,$ the figure compares the throughput in the cases with an ideal ($\epsilon=1, \vartheta=0, P_\text{RF}^\text{max}=\infty$) and non-ideal ($\epsilon=0.65, \vartheta=0.5, P_\text{RF}^\text{max}=18$ dB \cite{7104158}) PA. Then, Fig. 5 studies the effect of pointing error in the FSO link, and the outage probability is derived for different values of pointing error parameter $\xi.$ Here, the results are obtained for heterodyne detection technique and we set $ R=3 \text{ npcu}, N=100, \psi=0.25, \vartheta=0.5, \epsilon=0.65, P_\text{RF}^\text{max}=30 \text{ dB}$. Finally, Fig. 6 evaluates the outage probability for different data transmission techniques in the FSO link and coherence times of the RF and FSO links. The results of the figure are presented for $R=12 \text{ npcu}, M=1, \psi=2, \vartheta=0.5, \epsilon=0.65, P_\text{RF}^\text{max}=18 \text{ dB}, P=18 \text{ dB}, \xi=1.2,$ and the numbers of channel realizations $N$ for which central limit theorem provides accurate approximation for the random variable $\mathcal{Y}_{r,(m,N)}$ in (\ref{eq:eqWm1}). According to the results, the following conclusions can be drawn:
\begin{itemize}
  \item[1)] The analytical results of (\ref{eq:eqbessel0}) and (\ref{eq:eqasymrffso}) accurately mimic the exact numerical results, and the difference between the approximation-based and exact results is negligible for a broad range of SNRs/parameter settings (Fig. 3). Therefore, (\ref{eq:eqbessel0}) and (\ref{eq:eqasymrffso}) can effectively be used to investigate the performance of RF-FSO links analytically. Finally, note that in Figs. 4-6 the results are plotted based on exact evaluation of (\ref{eq:eqbessel0}). However, we have checked the results with the ones obtained via approximations (\ref{eq:eqbessel0}) and (\ref{eq:eqasymrffso}), and in all cases the approximation results are very tight.
  \item[2)] With different parameter settings, the implementation of HARQ leads to significant outage probability and energy efficiency improvement (Fig. 3). For instance, with the parameter settings of Fig. 3 and the outage probability $10^{-2},$ the implementation of HARQ with a maximum of $M=3$ retransmissions reduces the required power by almost $8$ dB, compared to the cases with open-loop communication ($M=1$). On the other hand, in harmony with the results on RF links \cite{6235077}, the HARQ may decrease the throughput of the RF-FSO links (Fig. 4). However, the throughput degradation is negligible for a broad range of parameter settings.
  %\item[3)] The outage probability decreases remarkably by the relative symbol rates of the RF and FSO links, and the effect of relative symbol rate parameter $\psi$ increases with the number of HARQ-based retransmissions $M$ (Fig. 4).
  \item[4)] The PAs inefficiency affects the performance of RF-FSO systems significantly (Fig. 4). For instance, with the parameter settings of Fig. 4, $M=1$ and $P=6$ dB, the inefficiency of the PA reduces the achievable throughput from $0.4$ npcu to $0.15$ npcu, i.e., $150\%$ throughput loss. However, the effect of imperfect PAs decreases with the SNR. This is intuitively because the \emph{effective} efficiency of the PAs $\epsilon^\text{effective}=\epsilon\left(\frac{P_\text{RF}}{P_\text{RF}^\text{max}}\right)^\vartheta$ increases with the SNR.
  \item[5)] The outage probability of the RF-FSO system is sensitive to severe pointing errors of the FSO link, i.e., small values of $\xi.$ However, the effect of pointing error is negligible for moderate/large values of $\xi,$ i.e., when the pointing error is negligible (Fig. 5). Also, with the parameter settings of Fig. 5, the outage probability achieved with severe pointing errors ($\xi=0.1$) and $M=2$ HARQ-based retransmission rounds is less than one in the open-loop setups ($M=1$) and no pointing error ($\xi\to\infty$). Thus, the HARQ can effectively be used to compensate the effect of pointing errors in RF-FSO links.
  \item[6)] As expected, better system performance is achieved by the implementation of heterodyne technique, compared to IM/DD technique (Fig. 6). Also, the outage probability decreases with increasing the number of channel realizations in the FSO link $N$ (Fig. 6). This is intuitively because more time diversity is exploited by the HARQ when the channel changes during the data transmission. Finally, in harmony with the intuitive understandings of (\ref{eq:eqasymrffso}), the system performance becomes insensitive to the number of channel realizations in the FSO link as $N$ increases.
\end{itemize}

\section{Conclusion}
In this paper, we studied the data transmission efficiency of mmwave-based RF-FSO links in the cases with and without HARQ feedback. Considering pointing errors in the FSO link and imperfect power amplifiers in the RF link, we derived closed-form expressions for the message decoding probabilities, throughput and outage probability of the RF-FSO systems. The results show that, while the throughput is not necessarily increasing by HARQ, substantial outage probability reduction is achieved by the implementation of HARQ protocols. Moreover, the inefficiency of the power amplifiers deteriorates the performance of RF-FSO systems considerably. Therefore, the properties of the power amplifiers should be carefully considered in the network design. Finally, the HARQ can be effectively utilized to compensate the effect of pointing errors in the FSO link.

\begin{figure}
\vspace{-0mm}
\centering
  \includegraphics[width=0.99\columnwidth]{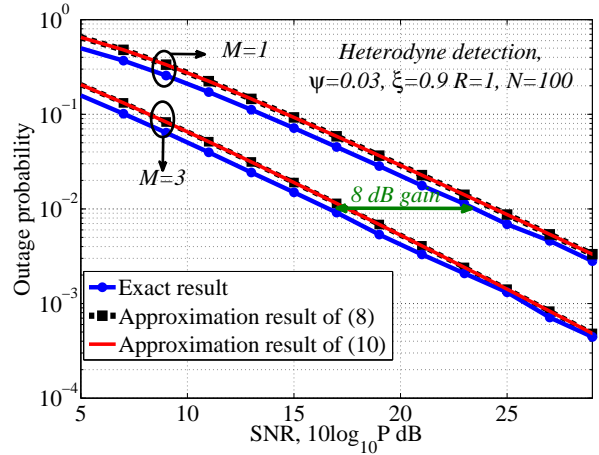}\\\vspace{-3mm}
\caption{On the tightness of the approximation results. Heterodyne detection technique, $ R=1 \text{ npcu}, N=100, \xi=0.9,$ and $ \psi=0.03$.  }\label{figure111}
\vspace{-3mm}
\end{figure}
%\begin{figure}
%\vspace{-0mm}
%\centering
%  % Requires \usepackage{graphicx}
%  \includegraphics[width=0.99\columnwidth]{figpointingerrorPsi.eps}\\\vspace{-0mm}
%\caption{Outage probability for different relative symbol rates of the RF and FSO links. IM/DD detection technique, $ R=3 \text{ npcu}, N=100,$ and $ \xi=0.9.$  }\label{figure111}
%\vspace{-0mm}
%\end{figure}
\vspace{-0mm}
\begin{figure}
\vspace{-0mm}
\centering
  \includegraphics[width=0.99\columnwidth]{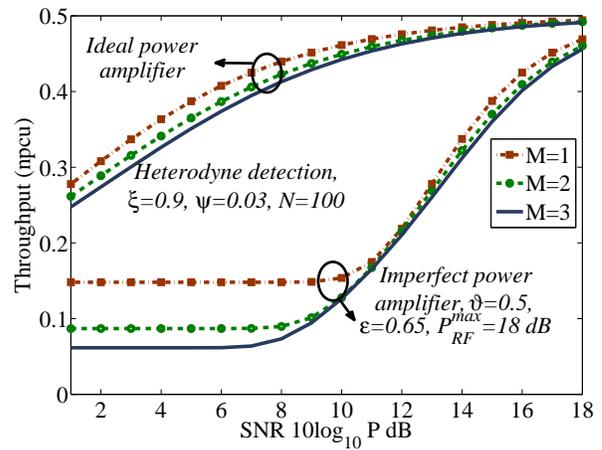}\\\vspace{-3mm}
\caption{Throughput for different types and power amplifiers and maximum numbers of retransmissions $M$. Heterodyne detection technique, $ R=0.5 \text{ npcu}, N=100, \xi=0.9, $ and $ \psi=0.03$.  }\label{figure111}
\vspace{-3mm}
\end{figure}
\begin{figure}
\vspace{-0mm}
\centering
  \includegraphics[width=0.99\columnwidth]{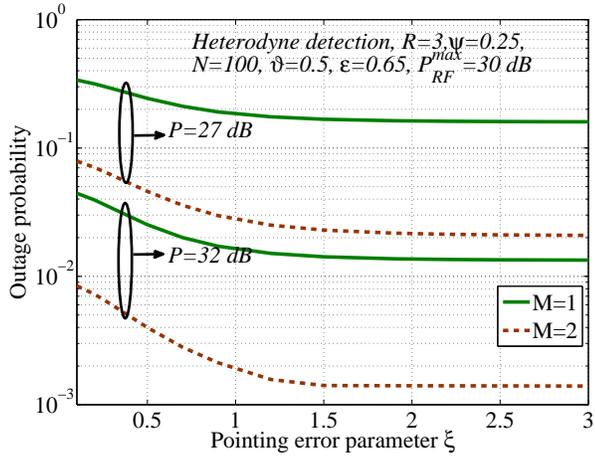}\\\vspace{-0mm}
\caption{Outage probability for different pointing errors in the FSO links. Heterodyne detection technique, $ R=3 \text{ npcu}, N=100, \psi=0.25, \vartheta=0.5, \epsilon=0.65,$ and $ P_\text{RF}^\text{max}=30 \text{ dB}$.  }\label{figure111}
\vspace{-0mm}
\end{figure}
\vspace{-0mm}
\begin{figure}
\vspace{-0mm}
\centering
  \includegraphics[width=0.99\columnwidth]{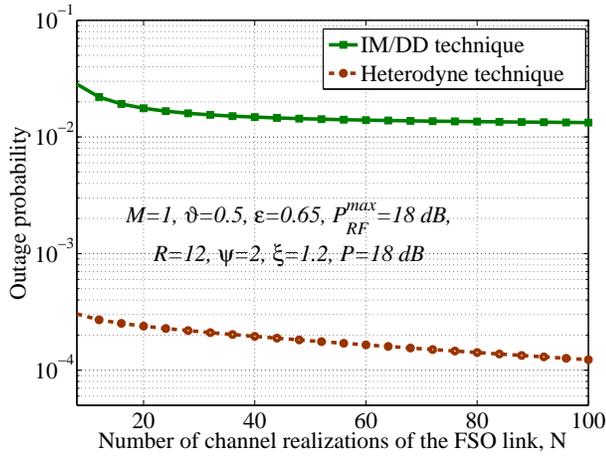}\\\vspace{-0mm}
\caption{Outage probability for different numbers of channel realizations of the FSO links $N$. The parameters are set to $ R=12 \text{ npcu}, M=1, \psi=2, \vartheta=0.5, \epsilon=0.65, P_\text{RF}^\text{max}=18 \text{ dB}, P=18 \text{ dB},$ and $\xi=1.2$.  }\label{figure111}
\vspace{-0mm}
\end{figure}
\vspace{-0mm}
\bibliographystyle{IEEEtran} %lic.bst is the style file
\bibliography{masterFSO3}
\vfill
\end{document}